\documentclass[a4paper]{jpconf}
\usepackage{graphicx}
\def\lsim{\mathrel{\rlap{
\lower4pt\hbox{\hskip-3pt$\sim$}}
    \raise1pt\hbox{$<$}}}     %less than approx. symbol
\def\gsim{\mathrel{\rlap{
\lower4pt\hbox{\hskip-3pt$\sim$}}
    \raise1pt\hbox{$>$}}}     %greater than or approx. symbol 
\begin{document}
\title{Baryon stopping signal for mixed phase formation in HIC}

\author{Yu. B. Ivanov}

\address{National Research Centre "Kurchatov Institute", 123182 Moscow, Russia}
\address{National Research Nuclear University "MEPhI" (Moscow Engineering
Physics Institute), 115409 Moscow, Russia}

\ead{y.b.ivanov@yandex.ru}

\begin{abstract}
It is argued that an irregularity in the baryon stopping is a natural consequence of 
onset of deconfinement 
occurring in the compression stage of a nuclear collision. It is an  effect 
of the softest point inherent in an equation of state (EoS) with 
a deconfinement transition.  
%Thus, this irregularity is a signal from a hot and dense stage of the nuclear collision.  
In order to illustrate this effect, calculations within the three-fluid model were performed 
with three different EoS's: a purely hadronic EoS, an EoS with a first-order phase transition  
and that with a smooth crossover transition.
\end{abstract}
%
%\today

\section{Introduction}

One of the main goals of the current experiments at RHIC and SPS and forthcoming
experiments at FAIR and NICA facilities is to determine a kind of the deconfinement
transition in dense  baryonic matter and to find the collision energy  
(thereby the baryonic density) at which this transition starts. 
In this paper it is argued that the baryon stopping 
in nuclear collisions can be a sensitive probe of the deconfinement onset. 
%In fact, an  irregularity 
%in the incident-energy dependence of the baryon stopping
% is very natural if the system undergoes a phase transition. 
Let us start with discussion in terms of the conventional 
(i.e. one-fluid) hydrodynamics. 
%if it is applied to the whole process of 
%the nuclear collision, e.g. from its compression stage
%to the expansion stage up to freeze-out.  

The form of the resulting rapidity distribution of net-baryons 
depends on the spatial form of the produced fireball. 
If the fireball is almost spherical, the expansion of the fireball is essentially 
3-dimensional which results in a peak at the midrapidity in the rapidity distribution. 
This statement is a theorem that can be proved in few lines. 
If a the fireball is strongly deformed (compressed) in the beam direction, 
i.e. has a form of a disk, its expansion is approximately 1-dimensional that 
produces a dip at the midrapidity, which is confirmed by numerous simulations, 
see e.g. \cite{71}. In terms of the fluid mechanics this is a consequence 
of interaction of two rarefaction waves propagating from opposite peripheral 
sides of decaying disk toward its center  \cite{Land-Lif}. 
%This speculation is very similar to that related to the elliptic flow: 
%a strong elliptic flow results from a strongly deformed almond-shaped 
%initial fireball with the  deformation of the resulting momentum 
%distribution of particles being inverse to the  spatial deformation of the 
%initial fireball. The physical mechanism here is precisely the same, only the 
%expansion is developed in the transverse direction.  

The formation of this fireball is already a matter of 
dynamics at the early compression stage of the nuclear collision. A softest point 
\cite{Hung:1994eq} characteristic of EoS's with a phase transition  
plays an important role in this compression dynamics. 
%The softest point in the equation of state is defined by a
%minimum in the ratio of the pressure to 
%the energy density at constant specific entropy (i.e. the entropy per baryon, $\sigma$): $(P/\varepsilon)_{\sigma}$. 
%The constancy of  is required because  it is conserved in the ideal 
%hydrodynamics. 
At the softest point the system exhibits the weakest resistance 
to its compression as compared with that in adjacent regions of the EoS. 
At low collision energies the softest point is not reached in the collision process, 
the system remains  stiff and therefore the produced fireball is 
almost spherical. As a result, the baryon rapidity distribution is
peaked at the midrapidity. 
When the incident energy gets high enough, the softest-point region  
of the EoS starts to dominate during the compression stage, the system weaker  
resists to the compression and hence the resulting fireball becomes more 
deformed, i.e. more of the disk shape. Then its expansion is close to the 
1-dimensional pattern and, as a result, we have a dip at the midrapidity. 
With energy rise, the stiffness of the EoS (in the range relevant to compression stage) 
grows, the system stats to be more resistant to the compression and hence 
the produced fireball becomes less deformed. The expansion of this 
fireball results in a peak or, at least, to a weaker dip at the midrapidity 
as compared to that at the ``softest-point'' incident energy. 
With further energy rise, the initial kinetic pressure overcomes 
the stiffness of the EoS and makes the produced fireball strongly deformed 
again, which in its turn again results in a dip at the midrapidity.

Thus, even without any nonequilibrium, we can 
expect a kind of a ``peak-dip-peak-dip'' irregularity in the 
incident energy dependence of the form of the net-proton rapidity distributions.
Nonequilibrium also contributes to this irregularity. 
At a phase transformation 
 the hadronic degrees of freedom are changed to partonic ones. 
In particular, the dip at the midrapidity in ultrarelativistic nuclear collisions 
occurs because the baryon charges of colliding nuclei traverse through 
each other rather than results from the 1-dimensional expansion of a disk-like fireball.

It is important to emphasize that the ``peak-dip-peak-dip'' irregularity 
is a signal from the hot and dense stage of the nuclear collision.

In the present paper this qualitative pattern is illustrated by calculations 
within a model of the three-fluid 
dynamics (3FD) \cite{3FD} employing three different equations of state (EoS): a purely hadronic EoS   
\cite{gasEOS} (hadr. EoS) and two versions of EoS involving  
deconfinement  \cite{Toneev06}:  
 an EoS with the first-order phase transition (2-phase EoS) 
and that with a smooth crossover transition (crossover EoS). The softest points in these EoS's 
 are illustrated in Ref. 
\cite{Nikonov:1998dg}. The hadronic EOS \cite{gasEOS} possesses no softest point, i.e.  
stiffness of the EoS changes monotonously. 
Results on the stopping power were reported in Refs. \cite{Ivanov:2010cu} in more ditail.

\section{Results of simulations}

A direct measure of the baryon stopping is the
net-baryon (i.e. baryons-minus-antibaryons) rapidity distribution. However, since experimental
information on neutrons is unavailable, we have to rely on net-proton (i.e. proton-minus-antiproton) data. 
Presently there exist experimental data on proton (or net-proton) rapidity spectra at 
AGS \cite{AGS} and 
SPS \cite{NA49-1} energies. 
%These data were analyzed within various models  
%Parton-Hadron-String Dynamics  (PHSD) \cite{Bratk09}, 
%Hybrid Hydro-Kinetic Model \cite{Bleicher09},
%Hadron-String Dynamics (HSD),  Ultra-relativistic Quantum Molecular Dynamics
%(UrQMD) \cite{Bratk04,WBCS03,Bratk02},
%Boltzmann-Uehling-Uhlenbeck (GiBUU) model 
%\cite{Larionov07,Larionov05} and three-fluid dynamics (3FD)
%\cite{3FD,3FD-GSI07}. 
%\cite{3FD,3FD-GSI07,Bratk09,Bleicher09,Bratk04,WBCS03,Bratk02,Larionov07,Larionov05}. 

\begin{figure}[bt]
\begin{center}
%\vspace*{-5mm}
%\hspace*{-14mm}
\includegraphics[width=12.9cm]{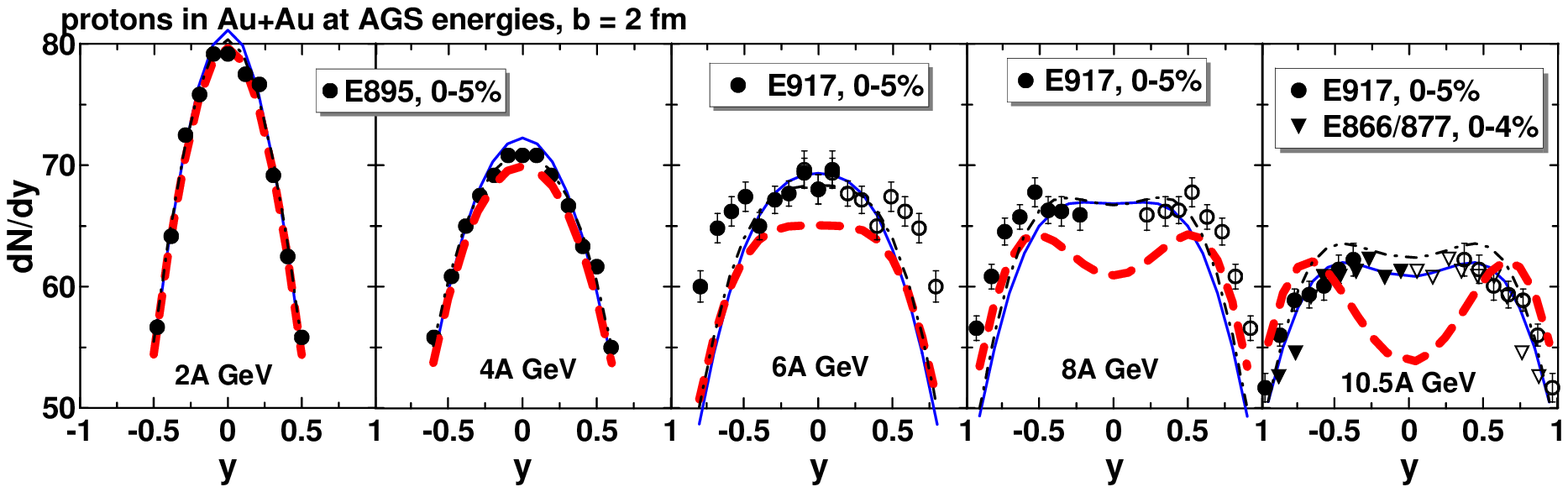}
\includegraphics[width=12.9cm]{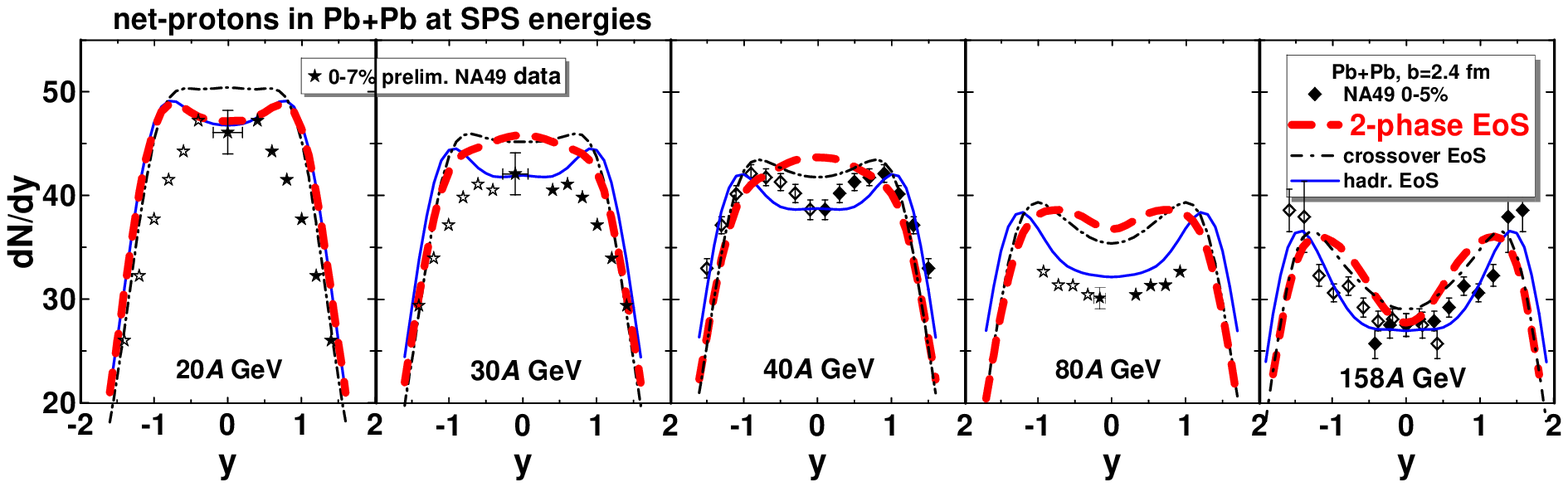}
\vspace*{-6mm}
\end{center}
 \caption{
Rapidity spectra of protons for AGS energies (upper raw of panel)
and net-protons SPS energies (lower raw of  panels) from  
central collisions of Au+Au (AGS) and Pb+Pb (SPS). 
Experimental data are from
\cite{AGS,NA49-1}. 
The percentage shows the fraction of the total reaction cross section, 
corresponding to experimental selection of central events. 
%Feedback of weak decays into yields is disregarded. 
} 
\label{fig4.1}
\end{figure}

Figure \ref{fig4.1} presents calculated rapidity distributions of net-protons 
in central collisions at AGS and SPS energies  
 and their comparison with available data.    
Difference between protons and net-protons is negligible at the AGS energies.   
As seen from Fig. \ref{fig4.1}, the distributions within the first-order-transition 
scenario indeed exhibit the above-discussed ``peak-dip-peak-dip'' irregularity in 
contrast to results obtained within the purely hadronic and crossover scenarios. 
The experimental distributions exhibit a qualitatively similar
behavior as that in the 2-phase-EoS scenario. 
However, quantitatively the 
2-phase-EoS results certainly disagree with data 
in the energy region  8$A$  GeV  $\leq E_{lab} \leq$ 40$A$  GeV.

\begin{figure}[tb]
%\vspace*{-54mm}
\begin{center}
\includegraphics[width=11.1cm]{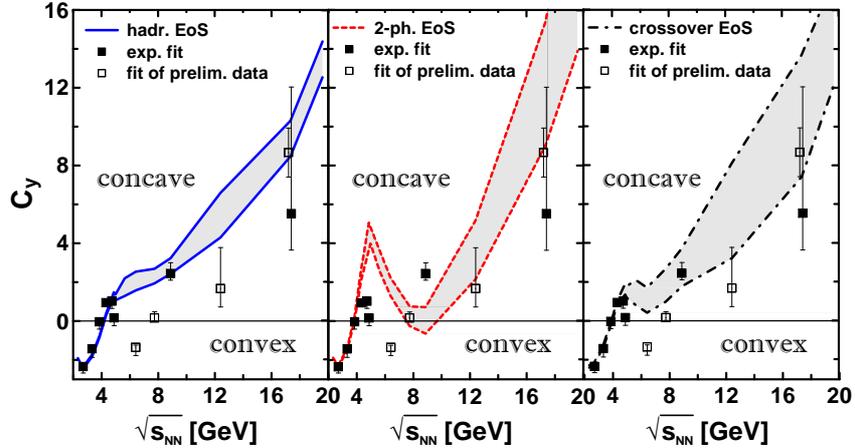}
%\vspace*{-59mm}
\end{center}
%\hspace*{6mm}
 \caption{\label{fig4a}
Midrapidity  reduced curvature  
   [see. Eq. (\ref{Cy})] 
of the (net)proton rapidity
   spectrum as a function of the center-of-mass energy
 of colliding nuclei as deduced from experimental data and predicted
 by 3FD calculations with  different EoS's: the hadronic EoS
   (hadr. EoS) \cite{gasEOS} (left panel),  the EoS involving a first-order phase
 transition (2-ph. EoS, middle panel) and the EoS with a crossover transition (crossover EoS, right panel) 
 into the quark-gluon phase \cite{Toneev06}. 
Upper bounds of the shaded areas correspond to fits confined in the region of $|y|/y_{\rm beam}<0.7$,  
lower bounds, $|y|/y_{\rm beam}<0.5$. 
}  
%\label{fig4a}
\end{figure}
In order to quantify the above-discussed ``peak-dip-peak-dip'' irregularity, it is useful to make use of the method proposed in Ref.~\cite{Ivanov:2010cu}. 
For this purpose the data on the net-proton rapidity distributions are fitted  by the following simple formula 
\begin{eqnarray}
\label{2-sources-fit} 
\frac{dN}{dy}=  
a \left(\exp\left\{ -(1/w_s)  \cosh(y-y_s) \right\}
%\right.
%\cr
+
%\left.
\exp\left\{-(1/w_s)  \cosh(y+y_s)\right\} \right)~,
\end{eqnarray}
where $a$, $y_s$ and $w_s$ are parameters of the fit. 
The form (\ref{2-sources-fit}) is a sum of two thermal sources shifted by $\pm
y_s$ from the midrapidity
which is put to be $y_{\rm mid}=0$ as it is in the collider mode. 
%
%The width $w_s$ of the sources can be interpreted as $w_s=$ (temperature)/(transverse mass), 
%if we assume that collective velocities in the sources have no spread with respect
%to the  source rapidities $\pm y_s$. 
The parameters of the two sources are identical (up to the sign of  $y_s$) because  only 
collisions of identical nuclei are considered. 
The above fit has been done by the least-squares method and 
applied to both available data and results of calculations.

A useful quantity, which characterizes the shape of the rapidity distribution, is
a reduced curvature of the spectrum at midrapidity defined as follows 
\begin{eqnarray}
\label{Cy} 
C_y = 
\left(y_{\rm beam}^3\frac{d^3N}{dy^3}\right)_{y=0}
\big/ \left(y_{\rm beam}\frac{dN}{dy}\right)_{y=0}
%\cr
=  
%\frac{2a y_{cm}^3}{w_s}\exp\left\{ -\frac{\cosh y_s}{w_s} \right\}
(y_{\rm beam}/w_s)^2 \left(
\sinh^2 y_s -w_s \cosh y_s 
\right), 
\end{eqnarray}
where $y_{\rm beam}$ is the beam rapidity in the collider mode.
 The second part of Eq. (\ref{Cy}) presents 
this curvature in terms of parameters of fit (\ref{2-sources-fit}).
 Excitation functions of $C_y$
deduced both from 
experimental data and from %the same fit (\ref{Cy}) of the 
results of the 3FD calculations 
with different EoS's are displayed in Fig. \ref{fig4a}. 
To evaluate errors of the $C_y$ values deduced from data, errors
produced by the least-squares method were estimated. 
The uncertainty associated with the choice of the 
rapidity range turned out to be the dominant one for the $C_y$
quantities deduced from simulation results. 
Therefore, in Fig.~\ref{fig4a} results for the curvature $C_y$
are presented by shaded areas with borders corresponding 
to the fit ranges $|y|<0.7 \; y_{\rm beam}$ and $|y|<0.5 \; y_{\rm beam}$.

The irregularity in data  is distinctly seen here 
as a wiggle irregularity in the energy dependence of $C_y$. 
Of course, this is only a hint to irregularity since this wiggle is formed only due to 
preliminary data of the NA49 collaboration. 
A remarkable observation is that 
the $C_y$ excitation function in the first-order-transition
scenario manifests qualitatively the same wiggle irregularity 
(middle panel of Fig. \ref{fig4a}) as
that in the data fit, while the hadronic  scenario produces purely monotonous 
behavior. 
The crossover EoS represents a very smooth transition, therefore, 
it is not surprising that it produces only a weak wiggle in $C_y$. 

\section{Conclusions}

In conclusion, 
the irregularity in the baryon stopping is a natural consequence of deconfinement  
occurring at the compression stage of a nuclear collision and thus  
is a signal from the hot and dense stage of the nuclear collision.
It is an effect of the softest point of a EoS.  
As it was demonstrated in Ref. \cite{Ivanov:2015vna}, this irregularity  is a very robust signal of a first-order phase transition that survives even under conditions of a very limited acceptance. 
Updated experimental results are badly needed 
to analyze a trend of the ``peak-dip-peak-dip'' irregularity. 
It would be highly desirable if new data are taken 
within the same experimental setup.
% and at the same centrality selection.  
%Hopefully such data will come from new accelerators FAIR and NICA. 

%\vspace*{2mm} {\bf Acknowledgements} \vspace*{1mm}

Fruitful collaboration with D. Blaschke
is gratefully acknowledged. 
I am grateful to A.S. Khvorostukhin, V.V. Skokov,  and V.D. Toneev for providing 
me with the tabulated 2-phase and crossover EoS's. 
The calculations were performed at the computer cluster of GSI (Darmstadt). 
This work was partially  supported  by  
%the Russian Ministry of Science and Education 
grant NS-215.2012.2.


\section*{References}
\begin{thebibliography}{999}


%%%%%%%%%%%%%%%%%%%%%%%%%%%%%%%%%%%%%%%%%%%%%%%%%%%%%%%%%%%%%%
% exp. data on proton rapidity distr.
%
\bibitem{71}
        Russkikh V N, Ivanov Y B, 
%        Dynamical Freeze-out in 3-Fluid Hydrodynamics,\\
     {\em Phys. Rev.} C {\bf 76} (2007) 054907. %  [nucl-th/0611094].
%
\bibitem{Land-Lif}
Landau L D and Lifshitz E M, {\em ``Fluid Mechanics''} (Pergamon
Press, Oxford, 1979).  
%%%%%%%%%%%%%%%%%%%%%%%%%%%%%%%%%%%%%%%%%%%%%%%%%%%%%%%%%%%%%%%%%%%%%%%%%
%'Softest Point'
%
\bibitem{Hung:1994eq} 
  Hung C~M~and~Shuryak E.~V.,
  %``Hydrodynamics near the QCD phase transition: Looking for the longest lived fireball,''
  {\em Phys.\ Rev.\ Lett.}\  {\bf 75} (1995) 4003. 
%  [hep-ph/9412360].
%%%%%%%%%%%%%%%%%%%%%%%%%%%%%%%%%%%%%%%%%%%%%%%%%%%%%%%%%%%%%%%%%%%%%%%%%
% 3FD Model
%
\bibitem{3FD}
 Ivanov Yu B, Russkikh V N, and Toneev V D,
 {\em Phys. Rev.} C {\bf 73} (2006) 044904. % [nucl-th/0503088].
%
\bibitem{gasEOS}
Galitsky V M and Mishustin I N, {\em Sov. J. Nucl. Phys.} {\bf 29}, 181
(1979).
%%%%%%%%%%%%%%%%%%%%%%%%%%%%%%%%%%%%%%%%%%%%%%%%%%%%%%%%%%%%%%%%%%%%%%%%
% Toneev EoS's
%
\bibitem{Toneev06}
 Khvorostukhin A S,  %{\it et al.},
 Skokov V V, Redlich K, and Toneev V D,
{\em Eur. Phys. J.} C {\bf 48} (2006) 531. % [nucl-th/0605069].
% 
%\cite{Nikonov:1998dg}
\bibitem{Nikonov:1998dg} 
~Nikonov  E~G,~Shanenko A~A and~Toneev V~D,
  %``A Mixed phase model and the 'Softest point' effect,''
  {\em Heavy Ion Phys.}\  {\bf 8} (1998) 89.
%  [nucl-th/9802018].
%
%%%%%%%%%%%%%%%%%%%%%%%%%%%%%%%%%%%%%%%%%%%%%%%%%%%%%%%%%%%%%%
% 3FD Model with ph. tr. 
%
\bibitem{Ivanov:2010cu}
Ivanov  Y~B,
  %``Baryon Stopping in Heavy-Ion Collisions at E_{lab}= 2--160 GeV/nucleon,''
  {\em Phys.\ Lett.}\  B {\bf 690} (2010) 358;
%  [arXiv:1001.0670 [nucl-th]].
%
%\cite{Ivanov:2011cb}
%\bibitem{Ivanov:2011cb} 
%  Y.~B.~Ivanov,
{\em Phys. At. Nucl.} {\bf 75} (2012) 621;  
  %``Baryon stopping as a signal of the mixed phase onset,''
%  [1101.2092 [nucl-th]].
%
%\bibitem{Ivanov:2012bh} 
%  Y.~B.~Ivanov,
  %``Baryon Stopping as a Probe of Deconfinement Onset in Relativistic Heavy-Ion Collisions,''
  {\em Phys.\ Lett.}\ B {\bf 721} (2013) 123;
%  [arXiv:1211.2579 [hep-ph]].
%
%\bibitem{Ivanov:2013wha} 
%  Y.~B.~Ivanov,
  %``Alternative Scenarios of Relativistic Heavy-Ion Collisions: I. Baryon Stopping,''
  {\em Phys.\ Rev.}\ C {\bf 87} (2013) 064904.
%  [arXiv:1302.5766 [nucl-th]].
%%%%%%%%%%%%%%%%%%%%%%%%%%%%%%%%%%%%%%%%%%%%%%%%%%%%%%%%%%%%%%
% exp. data
%
%\cite{AGS}
\bibitem{AGS} 
 Klay  J~L~{\em et al.}  [E-0895 Collaboration],
  %``Charged pion production in 2 to 8 agev central au+au collisions,''
  {\em Phys.\ Rev.}\ C {\bf 68} (2003) 054905;
%  [nucl-ex/0306033].
%\bibitem{E802} L.~Ahle {\em et al.} (E802 Collab.),
%Proton and deuteron production in Au+Au reactions at 11.6A GeV/c
%  Phys. Rev. C {\bf 60}, 064901 (1999).
%
%\bibitem{E877}  
Barrette J {\em et al.} (E877 Collaboration),
%Proton and pion production in Au+Au collisions at 10.8A GeV/c
{\em Phys. Rev.} C {\bf 62} (2000) 024901;
%
%\bibitem{E917} 
Back B B {\em et al.}, (E917 Collaboration),
% Baryon Rapidity Loss in Relativistic Au + Au Collisions
 {\em Phys. Rev. Lett.} {\bf 86 } (2001) 1970;
%
%\bibitem{E866} 
 Stachel J, {\em Nucl. Phys.} {\bf A654} (1999) 119c.
%[nucl-ex/9903007]. 
%
\bibitem{NA49-1} 
 Appelsh\"auser H {\em et al.} (NA49 Collaboration),
%Baryon Stopping and Charged Particle Distributions in Central Pb+Pb
%Collisions at 158 GeV per Nucleon
  {\em Phys. Rev. Lett.} {\bf 82} (1999) 2471;
%
%\bibitem{NA49-04}   
Anticic T {\em et al.} (NA49 Collaboration),
%Energy and centrality dependence of deuteron and proton production in
%Pb + Pb collisions at relativistic energies
   {\em Phys. Rev.} C {\bf 69} (2004) 024902;
%
%\bibitem{NA49-06}  
Alt C   {\em et al.} (NA49 Collaboration),
% Energy and centrality dependence of antiproton and proton production in relativistic Pb + Pb collisions at the CERN SPS.
% dN/dy in midrapidity
{\em Phys. Rev.} C {\bf 73} (2006) 044910;
%[nucl-ex/0512033].
%
%\bibitem{NA49-07}   
% p dN/dy at 7% select. 20A-80A GeV
Blume C (NA49 Collaboration), {\em J. Phys.} {\bf G34} (2007) S951;
%[nucl-ex/0701042].
%
%\bibitem{NA49-09}   
%\bibitem{Anticic:2010mp} 
  Anticic T.~{\em et al.}  [NA49 Collaboration],
  %``Centrality dependence of proton and antiproton spectra in Pb+Pb collisions at 40A GeV and 158A GeV measured at the CERN SPS,''
  {\em Phys.\ Rev.}\ C {\bf 83} (2011) 014901;
%  [arXiv:1009.1747 [nucl-ex]].
%exp. estimate of impact parameters
%\bibitem{Alt:2003ab} 
  Alt C~{\em et al.}  [NA49 Collaboration],
  %``Directed and elliptic flow of charged pions and protons in Pb + Pb collisions at 40-A-GeV and 158-A-GeV,''
  {\em Phys.\ Rev.}\ C {\bf 68} (2003) 034903.
%  [nucl-ex/0303001].
%%%%%%%%%%%%%%%%%%%%%%%%%%%%%%%%%%%%%%%%%%%%%%%%%%%%%%%%%%%%%
%\cite{Ivanov:2015vna}
\bibitem{Ivanov:2015vna} 
  Ivanov Y~B~and~Blaschke D,
  %``Robustness of the Baryon-Stopping Signal for the Onset of Deconfinement in Relativistic Heavy-Ion Collisions,''
    {\em Phys. Rev.} C {\bf 92} (2015) 024916. %  arXiv:1504.03992 [nucl-th].


\end{thebibliography}
\end{document}